\documentclass[italian,english]{article}
\usepackage[T1]{fontenc}
\usepackage[latin9]{inputenc}
\usepackage{amsmath}
\usepackage{amssymb}

\makeatletter
\newcommand{\lyxaddress}[1]{
\par {\raggedright #1
\vspace{1.4em}
\noindent\par}
}

\makeatother

\usepackage{babel}
\begin{document}

\title{\textbf{GRAVITATIONAL LUMINOSITY OF A HOT PLASMA IN $R^{2}$ GRAVITY }}

\author{\textbf{B. Nadiri Niri$^{1}$, A. Jahan$^{2}$, C. Corda$^{3}$}}
\maketitle

\lyxaddress{\textbf{$^{1,2,3}$Research Institute for Astronomy and Astrophysics
of Maragha (RIAAM), Maragha, IRAN}}

\lyxaddress{\textbf{Emails: $^{1}$bnbnadiri9@gmail.com, $^{2}$jahan@riaam.ac.ir,
$^{3}$cordac.galilei@gmail.com}}
\begin{abstract}
The $R^{2}$-gravity contribution to energy loss of a hot plasma due
to the gravitational bremsstrahlung is calculated in the linearized
theory on the basis of classical Coulomb scattering of plasma constituents
in small-angle scattering approximation. The explicit dependence of
the gravitational luminosity on the plasma temperature is derived
and its relevance to the Einstein gravity is demonstrated. The result
when applied to the Sun as a hot plasma, shows very good agreement
with available data.
\end{abstract}
\begin{description}
\item [{Keywords:}] \textbf{gravitational bremsstrahlung , $R^{2}$-gravity,
gravitational waves, small-angle scattering, hot plasma.}
\end{description}

\section{Introduction}

For a long time, the discovery of gravitational wave (GW) emissions
from the compact binary system with two neutron stars PSR1913+16 \cite{key-1}
has been the ultimate motivation for the design, implementation, and
advancement of extremely sophisticated GW detection technology. Physicists
working in this field of research need this technology to conduct
thorough investigations of GWs in order to advance science. The observation
of GWs from a binary black hole (BH) merger (event GW150914) \cite{key-2},
which occurred in the 100th anniversary of Albert Einstein's prediction
of GWs \cite{key-3}, has recently shown that this ambitious challenge
has been won. The event GW150914 represented a cornerstone for science
and for gravitational physics in particular. In fact, this remarkable
event equipped scientists with the means to give definitive proof
of the existence of GWs, the existence of BHs having mass greater
than 25 solar masses and the existence of binary systems of BHs which
coalesce in a time less than the age of the Universe \cite{key-2}.
After the event GW150914, LIGO detected a second burst of GWs from
merging BHs, the event GW151226 \cite{key-4}. The great hope is that
such detections, also through the collaboration with other detectors
\cite{key-5,key-6}, will soon become routine and part of a nascent
GW astronomy.

GW astronomy will be important for a better knowledge of the Universe
and also to confirm or to rule out the physical consistency of the
general theory of relativity (GTR) or of any other theory of gravitation
\cite{key-7}. This is because, in the context of extended theories
of gravity (ETG), some differences between the GTR and the others
theories can be pointed out starting by the linearized theory of gravity
\cite{key-7}. In this picture, detectors for GWs are in principle
sensitive also to a hypotetical \textit{scalar} component of gravitational
radiation, that appears in ETG like scalar-tensor gravity and $f(R)$
theories \cite{key-7}. Let us clarify some important motivations
which lead to a potential extension and generalization of the GTR. 

Although Einstein's GTR \cite{key-8} achieved great success (see
for example the opinion of Landau who says that the GTR is, together
with quantum field theory, the best scientific theory of all \cite{key-9})
and withstood many experimental tests, it also displayed many shortcomings
and flaws which today make theoreticians question whether it is the
definitive theory of gravity, see the reviews \cite{key-10,key-11,key-42}
and references within. As distinct from other field theories, like
the electromagnetic theory, the GTR is very difficult to quantize.
This fact rules out the possibility of treating gravitation like other
quantum theories, and precludes the unification of gravity with other
interactions. At the present time, it is not possible to realize a
consistent quantum theory of gravity (QTG) which leads to the unification
of gravitation with the other forces. From an historical point of
view, Einstein believed that, in the path to unification of theories,
quantum mechanics had to be subjected to a more general deterministic
theory, which he called generalized theory of gravitation, but he
did not obtain the final equations of such a theory (see for example
the biography of Einstein in \cite{key-12}). At present, this point
of view is partially retrieved by some theorists, starting from the
Nobel Laureate G. \textquoteright t Hooft \cite{key-13}. 

However, one has to recall that, during the last 30 years, a strong,
critical discussion about both the GTR and quantum mechanics has been
undertaken by theoreticians in the scientific community. The first
motivation for this historical discussion arises from the fact that
one of the most important goals of modern physics is to obtain a theory
which could, in principle, show the fundamental interactions as different
forms of the same \emph{symmetry} \cite{key-10,key-11,key-42}. Considering
this point of view, today one observes and tests the results of one
or more breaks of symmetry. In this way, it is possible to say that
we live in an \emph{unsymmetrical} world. In the last 60 years, the
dominant idea has been that a fundamental description of physical
interactions arises from quantum field theory. In this tapestry, different
states of a physical system are represented by vectors in a Hilbert
space defined in a spacetime, while physical fields are represented
by operators (i.e. linear transformations) on such a Hilbert space.
The greatest problem is that such a quantum mechanical framework is
not consistent with gravitation, because this particular field, i.e
the metric $h_{\mu\nu}$, describes both the dynamical aspects of
gravity and the spacetime background. In other words, one says that
the quantization of dynamical degrees of freedom of the gravitational
field is meant to give a quantum-mechanical description of the spacetime.
This is an unequalled problem in the context of quantum field theories,
because the other theories are founded on a fixed spacetime background,
which is treated like a classical continuum. Thus, at the present
time, an absolute QTG, which implies a total unification of various
interactions has not been obtained. In addition, the GTR assumes a
classical description of the matter which is totally inappropriate
at subatomic scales, which are the scales of the relic Universe \cite{key-14,key-15,key-42}.

In the unification approaches, from an initial point of view, one
assumes that the observed material fields arise from superstructures
like Higgs bosons or superstrings which, undergoing phase transitions,
generate actual particles. From another point of view, it is assumed
that geometry (for example the Ricci curvature scalar $R$) interacts
with material quantum fields generating back-reactions which modify
the gravitational action adding interaction terms (examples are high-order
terms in the Ricci scalar and/or in the Ricci tensor and non minimal
coupling between matter and gravity, see below). Various unification
approaches have been suggested, but without palpable observational
evidence in a laboratory environment on Earth. Instead, in cosmology,
some observational evidences could be achieved with a perturbation
approach \foreignlanguage{italian}{\cite{key-15,key-42}}. Starting
from these considerations, one can define as ETG those semi-classical
theories where the Lagrangian is modified, in respect of the standard
Einstein-Hilbert gravitational Lagrangian, adding high-order terms
in the curvature invariants (terms like $R^{2}$, $R^{\alpha\beta}R_{\alpha\beta}$,
$R^{\alpha\beta\gamma\delta}R_{\alpha\beta\gamma\delta}$, $R\Box R$,
$R\Box^{k}R$) or terms with scalar fields non-minimally coupled to
geometry (terms like $\phi^{2}R$), see \cite{key-10,key-11,key-42}
and references within. In general, one has to emphasize that terms
like those are present in all the approaches to the problem of unification
between gravity and other interactions. Additionally, from a cosmological
point of view, such modifications of the GTR generate inflationary
frameworks which are very important as they solve many problems of
the standard model of the Universe \cite{key-14,key-15,key-16,key-42}. 

In the general context of cosmological evidence, there are also other
considerations which suggest an extension of the GTR. As a matter
of fact, the accelerated expansion of the Universe, which is observed
today, implies that cosmological dynamics is dominated by the so called
Dark Energy, which gives a large negative pressure. This is the standard
picture, in which this new ingredient is considered as a source on
the right-hand side of the field equations. It should be some form
of un-clustered, non-zero vacuum energy which, together with the clustered
Dark Matter, drives the global dynamics. This is the so called \textquotedblleft concordance
model\textquotedblright{} ($\Lambda$CDM) which gives, in agreement
with the CMBR, LSS and SNeIa data, a good picture of the observed
Universe today, but presents several shortcomings such as the well
known \textquotedblleft coincidence\textquotedblright{} and \textquotedblleft cosmological
constant\textquotedblright{} problems \cite{key-17}. An alternative
approach is changing the left-hand side of the field equations, to
see if the observed cosmic dynamics can be achieved by extending the
GTR, see \cite{key-7,key-10,key-11,key-42} and references within.
In this different context, it is not required to find candidates for
Dark Energy and Dark Matter, that, till now, have not been found;
only the \textquotedblleft observed\textquotedblright{} ingredients,
which are curvature and baryon matter, have to be taken into account.
Considering this point of view, one can think that gravity is different
at various scales and there is room for alternative theories. In principle,
the most popular Dark Energy and Dark Matter models can be achieved
considering $f(R)$ theories of gravity, where $R$ is the Ricci curvature
\cite{key-7,key-10,key-11,key-42}. In this picture, the nascent GW
astronomy could, in principle, be important. In fact, a consistent
GW astronomy will be the definitive test for the GTR or, alternatively,
a strong endorsement for ETG \cite{key-7,key-42}. 

According to the GTR, a system with a time varying mass moment will
loss its energy by radiating the GWs \cite{key-3,key-9,key-18}. This
energy loss, at the lowest order, is proportional to the 3th order
time derivative of the quadrupole momentum of the mass-energy distribution
\cite{key-18}. In $R^{2}$-gravity, which is the simplest extension
of $f(R)$-gravity, because of the presence of third polarization
mode arising from the $R^{2}$ curvature term, the situation is different:
the extra massive mode contribution leads to an extra energy loss
which is proportional to 4th order time-derivative of the quadrupole
moment \cite{key-19}. 

By comparing the theoretical considerations with the observed decay
rate of binary systems PSR B1913+16 \cite{key-1} and PSR J0348+0432
\cite{key-20} some constraints on the strength of the $R^{2}$-dependent
term is obtained \cite{key-19,key-21,key-22}. In many astrophysical
situations, the hot plasma of ionized atoms emits electromagnetic
and gravitational radiation through the coulomb collisions between
the electrons and ions {[}23-26{]}. Thus, studying the gravitational
luminosity of a plasma is of general interest and may be another test
for the validity of $f(R)$ theory of gravity. In \cite{key-27} it
has been derived an expression for the amount of radiated energy in
a classical gravitational bremsstrahlung in $R^{2}$-gravity, assuming
the small-angle scattering approximation. In the present paper, we
apply it to derive the gravitational luminosity of a hot plasma with
the gravitational bremsstrahlung as a mechanism for the energy loss.
In Sec. 2, we linearize the $R^{2}$-gravity theory and, after that,
we briefly discuss the quadrupole radiation in $R^{2}$-gravity and
energy loss due to gravitational bremsstrahlung in a single Coulomb
collision between two charged particles. Sec. 3 is devoted to the
calculation of the thermal gravitational radiation of the hydrogen
plasma. In Sec. 4 we finally illustrate the correction with an application
to the Sun. A summary of the main results is presented in Sec. 5. 

\section{Linearized Theory and Quadrupole Radiation in Quadratic Gravity}

In the general framework of $f(R)$ gravity, the $\textbf{\ensuremath{R^{2}}}$
theory, which was originally proposed by Starobinski \cite{key-16},
has been analysed in various interesting works, see {[}28 - 32{]}
for example. Specifically, the non-singular behaviour of this class
of models is discussed in \cite{key-28}. In \cite{key-29} $\textbf{\ensuremath{R^{2}}}$
inflation is combined with the Dark Energy stage and in \cite{key-30}
an oscillating Universe, which is well tuned with some cosmological
observations is discussed. Finally, in \cite{key-31,key-32} the possibility
to partially solve the Dark Matter problem in the linearized $\textbf{\ensuremath{R^{2}}}$
theory has been analyzed.

It is also quite important to emphasize that the $\textbf{\ensuremath{R^{2}}}$
is the simplest one among the class of viable models with $R^{m}$
terms in addition to the Einstein-Hilbert theory. In \cite{key-33},
it has been shown that such models may lead to the (cosmological constant
or quintessence) acceleration of the universe as well as an early
time era of inflation. Moreover, they seem to pass the Solar System
tests, i.e. they have the acceptable newtonian limit, no instabilities
and no Brans-Dicke problem (decoupling of the scalar) in the scalar-tensor
version. 

The field equations of the $R^{2}$-gravity can be derived from the
action {[}28 - 33{]} 
\begin{equation}
I=\int d^{4}x\sqrt{-g}\big(a_{1}R+a_{2}R^{2}+16\pi G\mathcal{L}_{M}\big),\label{eq: action}
\end{equation}
where $\mathcal{L}_{M}$ is the Lagrangian density of matter and $a_{2}$
represents the coupling constant of the $R^{2}$ term. By varying
the action with respect to $g_{\mu\nu}$ one obtains 
\begin{equation}
G_{\mu\nu}+a_{2}\big(2R[R_{\mu\nu}-\frac{1}{2}g_{\mu\nu}R]-2R_{;\mu;\nu}+2g_{\mu\nu}\Box R\big)=T_{\mu\nu}^{(m)},\label{eq: R square field equation}
\end{equation}
with the associated Klein-Gordon equation obtained by taking the trace
of Eq. (\ref{eq: R square field equation}) as 
\begin{equation}
\Box R=E^{2}(R+T),\label{eq: Klein-Gordon}
\end{equation}
where $E$ is known as curvature energy term and defined via $E^{2}=\frac{1}{6a_{2}}$
\cite{key-32}. Relation (\ref{eq: Klein-Gordon}) implies the idea
of considering the Ricci scalar as an effective scalar field \cite{key-32}. 

Before starting the analysis, let us emphasize an important point.
As one wants the $R^{2}$-gravity theory to be viable, one needs that
it passes the Solar System tests. Thus, one must assume that the constant
coupling of the $\textbf{\ensuremath{R^{2}}}$ term in the gravitational
action results much minor in respect to the linear term $R\mbox{\mbox{. }}$
In this way, the variation from the standard GTR is very weak and
the theory can pass the Solar System tests. Regarding this important
issue, there are precedent works illustrating this and we need to
explicitly show that the bounds are respected. The key point is that
as the effective scalar field arising from curvature is very energetic,
then the constant coupling of the $\textbf{\ensuremath{R^{2}}}$ nonlinear
term $\rightarrow0$ \cite{key-34}. In this case, the Ricci curvature,
which is an extra dynamical quantity in the metric formalism, must
have a range longer than the size of the Solar System. An important
work is ref. \cite{key-35}, where it is shown that this is correct
if the effective length of the scalar field $l$ is much shorter than
the value of $0.2$ $mm$. In such a case, the presence of this effective
scalar is hidden from Solar System and terrestrial experiments. Another
important test concerns the deflection of light by the Sun. This effect
was studied in $\textbf{\ensuremath{R^{2}}}$ gravity by calculating
the Feynman amplitudes for photon scattering, and it was found that,
to linearized order, this deflection is the same as in the standard
GTR \cite{key-36}. In \cite{key-32} it has been shown that, in order
to partially solve the Dark Matter problem the value of the curvature
energy term implies a very low value of the constant coupling of the
$\textbf{\ensuremath{R^{2}}}$ term in the gravitational action, that
is $a_{2}\simeq10^{-34}cm^{4}$ in natural units. In that case, the
$R^{2}$-gravity theory results viable and $l\ll0.2$ $mm$ is guaranteed.

Now, let us proceed to linearize the $R^{2}$-gravity theory. We stress
that in the following linearization process we closely follow \cite{key-32}
with a small difference in the definition of the effective scalar
field. 

Starting frome eq. (\ref{eq: Klein-Gordon}), the identifications
\cite{key-37}

\begin{equation}
\begin{array}{ccccc}
\Phi\rightarrow2a_{2}R & +a_{1} & \textrm{and } &  & \frac{dV}{d\Phi}\rightarrow\frac{a_{1}R}{3}\end{array}\label{eq: identifica}
\end{equation}
permit to obtain a Klein - Gordon equation for the effective scalar
field $\Phi$ as

\begin{equation}
\square\Phi=\frac{dV}{d\Phi}.\label{eq: KG2}
\end{equation}
To study GWs, one analyzes the linearized theory in vacuum with a
little perturbation of the background, which is assumed given by a
a Minkowskian background plus $\Phi=\Phi_{0},$ that is, one linearizes
into a background with constant curvature \cite{key-32}. One also
assumes $\Phi_{0}$ to be a minimum for $V$ (natural units will be
used in the linearization process): 

\begin{equation}
V\simeq\frac{1}{2}\alpha\delta\Phi^{2}\Rightarrow\frac{dV}{d\Phi}\simeq m^{2}\delta\Phi,\label{eq: minimo}
\end{equation}
where the constant $m$ has mass dimension. Setting

\begin{equation}
\begin{array}{c}
g_{\mu\nu}=\eta_{\mu\nu}+h_{\mu\nu}\\
\\
\Phi=\Phi_{0}+\delta\Phi,
\end{array}\label{eq: linearizza}
\end{equation}
to first order in $h_{\mu\nu}$ and $\delta\Phi$, one calls $\widetilde{R}_{\mu\nu\rho\sigma}$
, $\widetilde{R}_{\mu\nu}$ and $\widetilde{R}$ the linearized quantity
which correspond to $R_{\mu\nu\rho\sigma}$ , $R_{\mu\nu}$ and $R$
\cite{key-32}. Thus, one writes down the linearized field equations
as \cite{key-32}:

\begin{equation}
\begin{array}{c}
\widetilde{R}_{\mu\nu}-\frac{\widetilde{R}}{2}\eta_{\mu\nu}=(\partial_{\mu}\partial_{\nu}h_{R}-\eta_{\mu\nu}\square h_{R})\\
\\
{}\square h_{R}=m^{2}h_{R},
\end{array}\label{eq: linearizzate1}
\end{equation}
with

\begin{equation}
h_{R}\equiv\frac{\delta\Phi}{\Phi_{0}}.\label{eq: definizione}
\end{equation}
$\widetilde{R}_{\mu\nu\rho\sigma}$ and eqs. (\ref{eq: linearizzate1})
are invariants for gauge transformations \cite{key-32}

\begin{equation}
\begin{array}{c}
h_{\mu\nu}\rightarrow h'_{\mu\nu}=h_{\mu\nu}-\partial_{(\mu}\epsilon_{\nu)}\\
\\
\delta\Phi\rightarrow\delta\Phi'=\delta\Phi.
\end{array}\label{eq: gauge}
\end{equation}
Thus, one defines \cite{key-32}

\begin{equation}
\bar{h}_{\mu\nu}\equiv h_{\mu\nu}-\frac{h}{2}\eta_{\mu\nu}+\eta_{\mu\nu}h_{R}.\label{eq: ridefiniz}
\end{equation}
Let us consider the transformation for the parameter $\epsilon^{\mu}$\cite{key-32}

\begin{equation}
\square\epsilon_{\nu}=\partial^{\mu}\bar{h}_{\mu\nu},\label{eq:lorentziana}
\end{equation}
which permits to choose a gauge analogous to the Lorenz one of electromagnetic
waves \cite{key-38}

\begin{equation}
\partial^{\mu}\bar{h}_{\mu\nu}=0.\label{eq: cond lorentz}
\end{equation}
Now, the field equations become \cite{key-32}

\begin{equation}
\square\bar{h}_{\mu\nu}=0\label{eq: onda T}
\end{equation}

\begin{equation}
\square h_{R}=m^{2}h_{R}\label{eq: onda S}
\end{equation}
The solutions of eqs. (\ref{eq: onda T}) and (\ref{eq: onda S})
are plan waves \cite{key-32}

\begin{equation}
\bar{h}_{\mu\nu}=A_{\mu\nu}(\overrightarrow{p})\exp(ip^{\alpha}x_{\alpha})+c.c.\label{eq: sol T}
\end{equation}

\begin{equation}
h_{R}=a(\overrightarrow{p})\exp(iq^{\alpha}x_{\alpha})+c.c.\label{eq: sol S}
\end{equation}
with

\begin{equation}
\begin{array}{ccc}
q^{\alpha}\equiv(\omega,\overrightarrow{p}) &  & \omega=p\equiv|\overrightarrow{p}|\\
\\
q^{\alpha}\equiv(\omega_{m},\overrightarrow{p}) &  & \omega_{m}=\sqrt{m^{2}+p^{2}}.
\end{array}\label{eq: k e q}
\end{equation}
Eqs. (\ref{eq: onda T}) and (\ref{eq: sol T}) represents the equation
and the solution for the standard tensor GWs of the GTR \cite{key-18}.
Eqs. (\ref{eq: onda S}) and (\ref{eq: sol S}) are respectively the
equation and the solution for the massive scalar mode instead \cite{key-32}.
We stress that the dispersion law for the modes of the massive scalar
field $h_{R}$ is not linear \cite{key-32}. In fact, the velocity
of the tensor modes $\bar{h}_{\mu\nu}$ is the light speed $c$, but
the dispersion law (the second of eq. (\ref{eq: k e q})) for the
modes of $h_{R}$ is that of a massive field which is interpreted
in terms of a wave-packet \cite{key-32}. We recall that the group-velocity
of a wave-packet of $h_{R}$ centered in $\overrightarrow{p}$ is
\cite{key-32}

\begin{equation}
\overrightarrow{v_{G}}=\frac{\overrightarrow{p}}{\omega_{m}}.\label{eq: velocita' di gruppo}
\end{equation}
This is exactly the velocity of a massive particle with mass $m$
and momentum $\overrightarrow{p}$. From the second of eqs. (\ref{eq: k e q})
and eq. (\ref{eq: velocita' di gruppo}) one gets

\begin{equation}
v_{G}=\frac{\sqrt{\omega_{m}^{2}-m^{2}}}{\omega_{m}}.\label{eq: velocita' di gruppo 2}
\end{equation}
As one wants a constant speed of the wave-packet, one obtains \cite{key-32}.

\begin{equation}
m=\sqrt{(1-v_{G}^{2})}\omega_{m}.\label{eq: relazione massa-frequenza}
\end{equation}
Let us continue our analysis in the Lorenz gauge \cite{key-38} with
trasformations of the type $\square\epsilon_{\nu}=0$; these trasformations
permit to obtain a condition of transversality for the tensor part
of the field: $k^{\mu}A_{\mu\nu}=0$ \cite{key-32}. On the other
hand, they do not give the transversality for the total field $h_{\mu\nu}$.
From eq. (\ref{eq: ridefiniz}) one gets \cite{key-32}

\begin{equation}
h_{\mu\nu}=\bar{h}_{\mu\nu}-\frac{\bar{h}}{2}\eta_{\mu\nu}+\eta_{\mu\nu}h_{R}.\label{eq: ridefiniz 2}
\end{equation}
At this point, if being in the massless case, one could set \cite{key-39}

\begin{equation}
\begin{array}{c}
\square\epsilon^{\mu}=0\\
\\
\partial_{\mu}\epsilon^{\mu}=-\frac{\bar{h}}{2}+h_{R}.
\end{array}\label{eq: gauge2}
\end{equation}
Eqs. (\ref{eq: gauge2}) give the total transversality of the field.
On the other hand, in the massive case this is impossible \cite{key-32}.
In fact, if one applies the Dalembertian operator to the second of
eqs. (\ref{eq: gauge2}) and uses the field equations (\ref{eq: onda T})
and (\ref{eq: onda S}), one gets \cite{key-32}

\begin{equation}
\square\epsilon^{\mu}=m^{2}h_{R},\label{eq: contrasto}
\end{equation}
which is in contrast with the first of eqs. (\ref{eq: gauge2}). In
the same way, it is possible to show that there is no linear relation
between the tensorial field $\bar{h}_{\mu\nu}$ and the massive scalar
field $h_{R}$ \cite{key-32}. Thus, one cannot choose a gauge in
which $h_{\mu\nu}$ is purely spatial (that is, one cannot set $h_{\mu0}=0,$
see eq. (\ref{eq: ridefiniz 2})) \cite{key-32}. One can set the
traceless condition to the field $\bar{h}_{\mu\nu}$ instead \cite{key-32}

\begin{equation}
\begin{array}{c}
\square\epsilon^{\mu}=0\\
\\
\partial_{\mu}\epsilon^{\mu}=-\frac{\bar{h}}{2}.
\end{array}\label{eq: gauge traceless}
\end{equation}
From eqs. (\ref{eq: gauge traceless}) one gets \cite{key-32}

\begin{equation}
\partial^{\mu}\bar{h}_{\mu\nu}=0.\label{eq: vincolo}
\end{equation}
If one wants to preserve the conditions $\partial_{\mu}\bar{h}^{\mu\nu}$
and $\bar{h}=0$ transformations like \cite{key-32}

\begin{equation}
\begin{array}{c}
\square\epsilon^{\mu}=0\\
\\
\partial_{\mu}\epsilon^{\mu}=0
\end{array}\label{eq: gauge 3}
\end{equation}
can be used. Thus, by taking $\overrightarrow{p}$ in the $z$ direction,
one chooses a gauge in which only $A_{11}$, $A_{22}$, and $A_{12}=A_{21}$
are different to zero \cite{key-32}. Setting $\bar{h}=0$ one gets
$A_{11}=-A_{22}$. Now, one puts these equations in eq. (\ref{eq: ridefiniz 2}),
obtaining

\begin{equation}
h_{\mu\nu}(t,z)=A^{+}(t-z)e_{\mu\nu}^{(+)}+A^{\times}(t-z)e_{\mu\nu}^{(\times)}+h_{R}(t-v_{G}z)\eta_{\mu\nu}.\label{eq: perturbazione totale}
\end{equation}
The term $A^{+}(t-z)e_{\mu\nu}^{(+)}+A^{\times}(t-z)e_{\mu\nu}^{(\times)}$
describes the two standard tensor GW polarizations which arise from
the GTR \cite{key-32}. The term $h_{R}(t-v_{G}z)\eta_{\mu\nu}$ is
the massive field arising from the $R^{2}$-gravity theory instead
\cite{key-32}. In other words, the Ricci scalar generates a third
massive GW polarization which is not present in the standard GTR \cite{key-32}.

Now, the post-newtonian expansion of the theory requires to assume
the space-time metric as a small perturbation expanded around the
flat background metric. In the following we restore CGS units. After
a lengthy algebra one finds the energy-momentum pseudo-tensor of the
gravitational field as \cite{key-19} 
\begin{equation}
t_{\mu\nu}=a_{1}k_{\mu}k_{\nu}\dot{h}_{\alpha\beta}\dot{h}^{\alpha\beta}-a_{2}\delta_{\mu\nu}(k_{\alpha}k_{\beta}\ddot{h}^{\alpha\beta})^{2}.\label{eq: pseudo-tensor}
\end{equation}
where $h_{\mu\nu}$ now denotes the fluctuating part of the space-time
metric, $k_{\mu}$ the 4-vector tangent to the world line of a GW
and $\dot{h}_{\alpha\beta}\equiv\partial_{0}h_{\alpha\beta}$. The
rate of energy loss of a matter system coupled to the gravity is found
to be \cite{key-9,key-18} 
\begin{eqnarray}
\frac{dE}{dt} & = & \int_{S}d\sigma\,\widehat{e}_{i}t^{0i}\nonumber \\
 & = & \frac{a_{1}}{60}G\dddot{Q}_{ij}\dddot{Q}^{ij}-\frac{a_{2}}{30}G\ddddot{Q}_{ij}\ddddot{Q}^{ij}.\label{eq: rate of energy loss}
\end{eqnarray}
The symbol $\widehat{e}_{i}$ stands for the unite vector along the
$i-th$ axis and the quadrupole moment of mass is defined to be \cite{key-9}
$Q_{ij}=mx_{i}x_{j}-r^{2}\delta_{ij}$. Some efforts are devoted to
determine the validity of above formula by probing the observational
parameters of the binary pulsar PSR 1913+16 \cite{key-19,key-21,key-22}.
Setting $a_{1}=\frac{4}{3}$ in the first term of above equation \cite{key-18},
re-produces the well-known energy loss of the GTR \cite{key-18}.
The gravitational energy radiated due to the Coulomb collision between
an electron with charge $e$ and speed $v$ and an ion with charge
$+Ze$, in small-angle scattering regime can be obtained as \cite{key-27}
\begin{eqnarray}
|\Delta E(b)| & = & G\int_{-\infty}^{\infty}dt\Big[\frac{a_{1}}{60}\dddot{Q}_{ij}(t)\dddot{Q}^{ij}(t)-\frac{a_{2}}{30}G\ddddot{Q}_{ij}\ddddot{Q}^{ij}\Big],\nonumber \\
 & = & \frac{1}{24}\frac{Z^{2}e^{4}v\pi G}{b^{3}}a_{1}-B,\label{eq: energy radiated}
\end{eqnarray}
where $B\equiv\frac{a_{2}}{30}\ddddot{Q}_{ij}(t)\ddddot{Q}^{ij}(t)$
and $b$ denotes the impact parameter. In small-angle approximation,
one considers the particle's trajectory as a straight line \cite{key-25,key-26}.
Let us compute $B,$ that is the contribution of the $R^{2}$ term
to the gravitational energy loss. For the time derivatives of the
quadruple moment one gets 
\begin{eqnarray}
\dddot{Q}_{ij} & = & \mu\big[3\dddot{x}_{i}x_{j}+9\ddot{x}_{i}\dot{x}_{j}+9\dot{x}_{i}\ddot{x}_{j}+3x_{i}\dddot{x}_{j}-2(3\dot{\textbf{x}}\cdot\ddot{\textbf{x}}+{\textbf{x}}\cdot\dddot{\textbf{x}})\delta_{ij}\big]\\
\ddddot{Q}_{ij} & = & \mu\big[3\ddddot{x}_{i}x_{j}+12\dddot{x}_{i}\dot{x}_{j}+18\ddot{x}_{i}\ddot{x}_{j}+12\dot{x}_{i}\dddot{x}_{j}+3x_{i}\ddddot{x}_{j}\\
 &  & -2(3\ddot{\textbf{x}}\cdot\ddot{\textbf{x}}+4\dot{\textbf{x}}\cdot\dddot{\textbf{x}}+\textbf{x}\cdot\ddddot{\textbf{x}})\delta_{ij}\big]\nonumber 
\end{eqnarray}
Thus, one has 
\begin{equation}
\begin{array}{c}
\dddot{Q}_{ij}\dddot{Q}^{ij}=2\ m^{2}\big(3\dddot{\textbf{x}}\cdot\dddot{\textbf{x}}\textbf{x}\cdot\textbf{x}+18\dddot{\textbf{x}}\cdot\ddot{\textbf{x}}\dot{\textbf{x}}\cdot\textbf{x}+18\dddot{\textbf{x}}\cdot\dot{\textbf{x}}\ddot{\textbf{x}}\cdot\textbf{x}+\dddot{\textbf{x}}\cdot{\textbf{x}}\dddot{\textbf{x}}\cdot\textbf{x}\\
\\
+27\ddot{\textbf{x}}\cdot\ddot{\textbf{x}}\dot{\textbf{x}}\cdot\dot{\textbf{x}}+9\ddot{\textbf{x}}\cdot\dot{\textbf{x}}\ddot{\textbf{x}}\cdot\dot{\textbf{x}}-12\dddot{\textbf{x}}\cdot{\textbf{x}}\ddot{\textbf{x}}\cdot\dot{\textbf{x}}\big).
\end{array}
\end{equation}
Then, 
\begin{equation}
\begin{array}{c}
\ddddot{Q}_{ij}\ddddot{Q}^{ij}=2\ m^{2}\big(3\ddddot{\textbf{x}}\cdot\ddddot{\textbf{x}}\textbf{x}\cdot\textbf{x}+24\ddddot{\textbf{x}}\cdot\dddot{\textbf{x}}\dot{\textbf{x}}\cdot\textbf{x}+36\ddddot{\textbf{x}}\cdot\ddot{\textbf{x}}\ddot{\textbf{x}}\cdot\textbf{x}+24\ddddot{\textbf{x}}\cdot\dot{\textbf{x}}\dddot{\textbf{x}}\cdot\textbf{x}\\
\\
+\ddddot{\textbf{x}}\cdot\textbf{x}\ddddot{\textbf{x}}\cdot\textbf{x}-12\ddddot{\textbf{x}}\cdot\textbf{x}\ddot{\textbf{x}}\cdot\ddot{\textbf{x}}-16\ddddot{\textbf{x}}\cdot\textbf{x}\dddot{\textbf{x}}\cdot\dot{\textbf{x}}+144\dddot{\textbf{x}}\cdot\ddot{\textbf{x}}\ddot{\textbf{x}}\cdot\dot{\textbf{x}}\\
\\
+16\dddot{\textbf{x}}\cdot\dot{\textbf{x}}\dddot{\textbf{x}}\cdot\dot{\textbf{x}}-48\dddot{\textbf{x}}\cdot\dot{\textbf{x}}\ddot{\textbf{x}}\cdot\ddot{\textbf{x}}+36\ddot{\textbf{x}}\cdot\ddot{\textbf{x}}\ddot{\textbf{x}}\cdot\ddot{\textbf{x}}+48\dot{\textbf{x}}\cdot\dot{\textbf{x}}\dddot{\textbf{x}}\cdot\dddot{\textbf{x}}\big).
\end{array}
\end{equation}
Using 
\begin{eqnarray}
\ddot{\mathbf{x}} & = & \frac{\gamma vt}{r^{3}}\widehat{e}_{x}+\frac{\gamma b}{r^{3}}\widehat{e}_{y}\\
\dddot{\mathbf{x}} & = & \gamma\Big(\frac{v}{r^{3}}-3\frac{v^{3}t^{2}}{r^{5}}\Big)\widehat{e}_{x}-3\frac{\gamma v^{2}tb}{r^{5}}\widehat{e}_{y}\\
\ddddot{\mathbf{x}} & = & \gamma\Big(15\frac{v^{5}t^{3}}{r^{7}}-9\frac{v^{3}t}{r^{5}}-2\frac{\gamma vt}{r^{6}}\Big)\widehat{e}_{x}+\gamma\big(15\frac{v^{4}t^{2}b}{r^{7}}-3\frac{v^{2}b}{r^{5}}-2\frac{\gamma b}{r^{6}}\Big)\widehat{e}_{y}
\end{eqnarray}
with $\ensuremath{\gamma=\frac{Ze^{2}}{m_{e}}},$ from (36-38) one
constructs the following set of relations 
\begin{eqnarray}
\ddddot{\mathbf{x}}\cdot\ddddot{\mathbf{x}} & = & 45\frac{\gamma^{2}v^{8}t^{4}}{r^{12}}-18\frac{\gamma^{2}v^{6}t^{2}}{r^{10}}+9\frac{\gamma^{2}v^{4}}{r^{8}}+O(\gamma^{3})\\
\ddddot{\mathbf{x}}\cdot\dddot{\mathbf{x}} & = & -12\frac{\gamma^{2}v^{6}t^{2}}{r^{10}}-2\frac{\gamma^{2}v^{4}t}{r^{8}}+O(\gamma^{3})\\
\ddddot{\mathbf{x}}\cdot\ddot{\mathbf{x}} & = & 9\frac{\gamma^{2}v^{4}t^{2}}{r^{8}}-3\frac{\gamma^{2}v^{2}}{r^{6}}+O(\gamma^{3})\\
\ddddot{\mathbf{x}}\cdot\dot{\mathbf{x}} & = & 15\frac{\gamma v^{6}t^{3}}{r^{7}}-9\frac{\gamma v^{4}t}{r^{5}}-2\frac{\gamma^{2}v^{2}t}{r^{6}}\\
\ddddot{\mathbf{x}}\cdot{\mathbf{x}} & = & 9\frac{\gamma v^{4}t^{2}}{r^{5}}-2\frac{\gamma^{2}}{r^{4}}-3\frac{\gamma v^{2}}{r^{3}}\\
\dddot{\mathbf{x}}\cdot\dddot{\mathbf{x}} & = & 3\frac{\gamma^{2}v^{4}t^{2}}{r^{8}}+\frac{\gamma^{2}v^{2}}{r^{6}}\\
\dddot{\mathbf{x}}\cdot\ddot{\mathbf{x}} & = & -2\frac{\gamma^{2}v^{2}t}{r^{6}}\\
\dddot{\mathbf{x}}\cdot\dot{\mathbf{x}} & = & -3\frac{\gamma v^{4}t^{2}}{r^{5}}+\frac{\gamma v^{2}}{r^{3}}.
\end{eqnarray}
By the help of (35), for the time derivatives of the quadrupole moment
one obtains 
\begin{eqnarray}
\ddddot{Q}_{ij}\ddddot{Q}^{ij} & = & 2\mu^{2}\Big(1224\frac{\gamma^{2}v^{8}t^{4}}{r^{10}}-612\frac{\gamma^{2}v^{6}t^{2}}{r^{8}}+132\frac{\gamma^{2}v^{4}}{r^{6}}-288\frac{\gamma^{2}v^{8}t^{3}}{r^{10}}\\
 &  & -1080\frac{\gamma^{2}v^{10}t^{6}}{r^{12}}-1080\frac{\gamma^{2}v^{8}t^{4}b^{2}}{r^{12}}+648\frac{\gamma^{2}v^{6}t^{2}b^{2}}{r^{10}}\Big).\nonumber 
\end{eqnarray}
Therefore, one gets the contribution of the $R^{2}$ term to the gravitational
energy loss as 
\begin{eqnarray}
B & = & G\int_{-\infty}^{\infty}\frac{a_{2}}{30}\ddddot{Q}_{ij}(t)\ddddot{Q}^{ij}(t)dt=\\
 &  & =-\frac{a_{2}}{15}Gm^{2}\int_{-\infty}^{\infty}dt\Big[1224\frac{\gamma^{2}v^{8}t^{4}}{(v^{2}t^{2}+b^{2})^{5}}-612\frac{\gamma^{2}v^{6}t^{2}}{(v^{2}t^{2}+b^{2})^{4}}+132\frac{\gamma^{2}v^{4}}{(v^{2}t^{2}+b^{2})^{3}}+\nonumber \\
 &  & -288\frac{\gamma^{2}v^{8}t^{3}}{(v^{2}t^{2}+b^{2})^{5}}-1080\frac{\gamma^{2}v^{10}t^{6}}{(v^{2}t^{2}+b^{2})^{6}}-1080\frac{\gamma^{2}v^{8}t^{4}b^{2}}{(v^{2}t^{2}+b^{2})^{6}}+648\frac{\gamma^{2}v^{6}t^{2}b^{2}}{(v^{2}t^{2}+b^{2})^{5}}\Big]\nonumber \\
 &  & =\frac{213}{80}\frac{Z^{2}e^{4}v^{3}\pi G}{b^{5}}a_{2}.\nonumber 
\end{eqnarray}
The set of the integrals that have been used in evaluating (48) can
be found in the Appendix.

Thus, eq. (\ref{eq: energy radiated}) becomes 
\begin{eqnarray}
|\Delta E(b)| & = & G\int_{-\infty}^{\infty}dt\Big[\frac{a_{1}}{60}\dddot{Q}_{ij}(t)\dddot{Q}^{ij}(t)-\frac{a_{2}}{30}G\ddddot{Q}_{ij}\ddddot{Q}^{ij}\Big],\nonumber \\
 & = & \frac{1}{24}\frac{Z^{2}e^{4}v\pi G}{b^{3}}a_{1}-\frac{213}{80}\frac{Z^{2}e^{4}v^{3}\pi G}{b^{5}}a_{2},\label{eq: energy radiated-1}
\end{eqnarray}

\section{Thermal Gravitational Radiation of a Hot Plasma }

To obtain the gravitational luminosity of a plasma with the gravitational
bremsstrahlung as a mechanism for the loss of its energy, one must
multiply (\ref{eq: energy radiated}) with the electron flux $vn_{e}$,
ion density $n_{i}$, and integrate over the impact parameter $b$
\cite{key-25,key-26}. Therefore, we obtain the luminosity $\mathcal{L}$
(energy loss per volume $V$) of the plasma as 
\begin{equation}
\mathcal{L}=\frac{d\mathcal{E}}{dV}=2\pi n_{i}n_{e}v\int_{b_{min}}^{\infty}db|\Delta E(b)|b.\label{eq: luminosity}
\end{equation}
This integral diverges as $b\rightarrow0$. Thus a cut-off, denoted
by $b_{min}$, is introduced to get a finite result for the luminosity.
Based on either classical or quantum mechanical considerations the
cut-off takes the form \cite{key-25,key-26} 
\begin{equation}
b_{min}=\left\{ \begin{array}{ll}
\frac{e^{2}}{m_{e}v^{2}}\\
\\
\frac{\hbar}{m_{e}v}
\end{array}\right.
\end{equation}
respectively. The final result for the luminosity depends on which
form for the cut-off is engaged. We will restrict ourself to the Hydrogen
plasma. Thus, $Z=1$ and $n_{e}=n_{i}$. Hence from (\ref{eq: luminosity}),
with quantum mechanical cut-off $b_{min}=\frac{\hbar}{m_{e}v}$, the
energy loss takes the form 
\begin{equation}
\mathcal{L}=2\pi^{2}\frac{e^{4}n_{e}^{2}G}{c^{5}}\bigg(\frac{1}{24}\frac{m_{e}v^{3}}{\hbar}a_{1}-\frac{213}{240}\frac{m_{e}^{3}v^{7}}{\hbar^{3}}a_{2}\bigg).\label{eq: energy loss}
\end{equation}
One notes that the speed of light, $c$ is restored in (\ref{eq: energy loss}).

By taking thermal average of the above expression, one gets the thermal
luminosity of the plasma. In many astrophysical objects, the ratio
of Coulomb interaction energy to thermal energy is negligible, so
the hot plasma behaves like an almost ideal gas \cite{key-26}. Thus,
one can calculate the thermal luminosity of the plasma by averaging
the electron speed in (\ref{eq: energy loss}) over a thermal distribution
of speeds. For an ensemble of particles at temperature $T$, obeying
the Maxwell-Boltzman statistic, the thermal average is 
\begin{equation}
\langle f(\mathbf{v})\rangle=\bigg(\frac{m\beta}{2\pi}\bigg)^{\frac{3}{2}}\int d^{3}ve^{-\frac{\beta}{2}mv^{2}}f(\mathbf{v}),\quad\beta=\frac{1}{k_{B}T}.\label{eq: thermal average}
\end{equation}
where $f(\mathbf{v})$ is an arbitrary function of particle's velocity
and $k_{B}$ denotes the Boltzman constant. In particular one obtains
\begin{eqnarray}
\langle v^{2n+1}\rangle & = & \frac{2}{\pi^{n+1}}\bigg(\frac{2\pi}{\beta m}\bigg)^{n+\frac{1}{2}}(n+1)!,\quad\quad(0\leq n)\label{eq: 36}\\
\langle v^{2n}\rangle & = & \frac{(2n+1)!!}{\beta^{n}m^{n}}.\label{eq: 37}
\end{eqnarray}
Thus, from (\ref{eq: energy loss}) and (54) we obtain the gravitational
luminosity with the quantum mechanical cut-off as 
\begin{equation}
\begin{array}{c}
\langle\mathcal{L}\rangle=2\pi^{2}\frac{e^{4}n_{e}^{2}G}{c^{5}}\bigg(\frac{1}{12}\frac{m_{e}a_{1}}{\hbar}\langle v^{3}\rangle-\frac{213}{120}\frac{m_{e}^{3}a_{2}}{\hbar^{3}}\langle v^{7}\rangle\bigg)=\\
\\
=\sqrt{2\pi^{3}}\frac{e^{4}n_{e}^{2}G}{c^{5}}\bigg[\frac{4}{6}\frac{m_{e}a_{1}}{\hbar}\bigg(\frac{k_{B}T}{m_{e}}\bigg)^{\frac{3}{2}}-\frac{3408}{5}\frac{m_{e}^{3}a_{2}}{\hbar^{3}}\bigg(\frac{k_{B}T}{m_{e}}\bigg)^{\frac{7}{2}}\bigg].
\end{array}\label{eq: 38}
\end{equation}
Now, it is evident how the presence of the $a_{2}R^{2}$ term in the
action affects the gravitational luminosity of a hot plasma. Equation
(\ref{eq: 38}) stands as our final result for the gravitational luminosity.
Setting $a_{2}=0$ in (\ref{eq: 38}) yields 
\begin{equation}
\langle\mathcal{L}\rangle\sim\frac{e^{4}n_{e}^{2}Gm_{e}}{c^{5}\hbar}\bigg(\frac{k_{B}T}{m_{e}}\bigg)^{\frac{3}{2}}.\label{eq: 39}
\end{equation}
which is the well-known result derived earlier by Weinberg within
the context of the GTR \cite{key-23}

\section{Gravitational Luminosity of the Sun }

For an astrophysical application we use Eq. (\ref{eq: 38}) to calculate
the gravitational energy loss of the Sun within the framework of the
$R^{2}$-gravity theory. The total gravitational luminosity of the
Sun is $L_{\odot}=V_{\odot}\langle\mathcal{L}_{\odot}\rangle$ where
$V_{\odot}$ denotes the Sun's volume. By setting $a_{1}=\frac{4}{3}$,
Eq. (\ref{eq: 38}) takes the form
\begin{equation}
\begin{array}{c}
L_{\odot}=L_{\odot}^{(1)}+L_{\odot}^{(2)}=\\
\\
\left[\sqrt{2\pi^{3}}\,\frac{8}{9}\frac{m_{e}e^{4}n_{e}^{2}G}{\hbar c^{5}}\bigg(\frac{k_{B}T}{m_{e}}\bigg)^{\frac{3}{2}}-\frac{3408}{5}\sqrt{2\pi^{3}}\,a_{2}\frac{m_{e}^{3}e^{4}n_{e}^{2}G}{c^{5}\hbar^{3}}\bigg(\frac{k_{B}T}{m_{e}}\bigg)^{\frac{7}{2}}\right]V_{\odot}.
\end{array}
\end{equation}
Based on the massive scalar mode arising from the $R^{2}$ term, the
coupling constant $a_{2}$ comes to be very small with respect to
linear term $R$. Assuming the typical galactic scale for the curvature
energy, $E\simeq10^{45}g$, we find $a_{2}=10^{-34}cm^{4}$ in natural
units \cite{key-32}. In this way, the variation from the standard
GTR is very weak. The parameters needed to obtain the above result
(in CGS units) are 
\begin{eqnarray}
m_{e} & = & 9\times10^{-28}\textrm{gr}\\
e & = & 4.8\times10^{-10}\textrm{esu}\\
G & = & 6.67\times10^{-8}\textrm{cm}^{3}.\textrm{gr}^{-1}.\textrm{sec}^{-2}\\
\hbar & = & 10^{-27}\textrm{erg.sec}\\
k_{B} & = & 1.38\times10^{-16}\textrm{erg.K}^{-1}\\
c & = & 3\times10^{10}\textrm{cm.sec}^{-1}\\
n_{e} & = & 3\times10^{25}\textrm{cm}^{-3}\\
V_{\odot} & = & 2\times10^{31}\textrm{cm}^{3}\\
T_{\odot} & = & 10^{7}\textrm{K}
\end{eqnarray}
Therefore one straightforwardly calculates 
\begin{eqnarray}
L_{\odot}^{(1)} & \simeq & 10^{16}\textrm{erg.sec}^{-1}\\
L_{\odot}^{(2)} & \simeq & 10^{13}\textrm{erg.sec}^{-1}
\end{eqnarray}
As we can see the first term in Eq. (68) that coming from the standard
GTR is in good agreement with available results \cite{key-23,key-24,key-40}.
It is important to clarify the physical reason of the very small contribution
of the $R^{2}$ term in Eq. (69). The reader could indeed think that
we did not take into account all effects of the $R^{2}$ gravity theory
under investigation or that we did not correctly chose the PPN-restricted
parameters for the action (1). In fact, there maybe GWs enhancing
from some models of f(R) gravity of the order of 15\% \cite{key-41}.
The key point here is that, exactly in order to match the PPN-restricted
parameters for the action (1) and to be consistent with solar system
tests, we have set the coupling constant of the $R^{2}$ term very
small. Such a setting has been chosen also to match the Dark Matter
model in \cite{key-31,key-32}. This crucial point makes the contribution
of Eq. (69) small.

\section{Concluding remarks }

A new era in astrophysics and gravitation started with the events
GW150914 \cite{key-2} and GW151226 \cite{key-4}. In fact, on one
hand the nascent GW astronomy will be important for a better knowledge
of the Universe. On the other hand, it will permit to confirm or to
rule out the physical consistency of the GTR or of any other theory
of gravitation \cite{key-7}. A key point is indeed that, in the framework
of the ETG, some differences between the GTR and the others theories
can be pointed out starting by the linearized theory of gravity \cite{key-7}.
Some important motivations which lead to a potential extension and
generalization of the GTR have been stressed in the Introduction of
this paper. The most important issue is, perhaps, the possibility
to see the ETG as a potential alternative to Dark Matter and Dark
Energy \cite{key-7,key-10,key-11}. Considering this different approach,
gravity could be different at different scales and there is room for
alternative theories. In fact, Dark Energy and Dark Matter can be,
in principle, achieved if one considers $f(R)$ theories of gravity,
where $R$ is the Ricci curvature \cite{key-7,key-10,key-11}. In
this alternative framework, the nascent GW astronomy should be important
becaues a consistent GW astronomy will be the definitive test for
the GTR or, alternatively, a strong endorsement for ETG \cite{key-7}. 

In the GTR, a system with a time varying mass moment will loss its
energy by radiating gravitational radiation \cite{key-3,key-9,key-18}.
At the lowest order the energy loss is proportional to the 3th order
time derivative of the quadrupole momentum of the mass-energy distribution
\cite{key-18}. $R^{2}$-gravity represents the simplest extension
of $f(R)$-gravity. It shows the presence of a third polarization
mode arising from the $R^{2}$ curvature term. In that case, the situation
is different. In fact, the extra massive mode contribution leads to
an extra energy loss which is proportional to 4th order time-derivative
of the quadrupole moment \cite{key-19}. If one compares the theoretical
considerations with the observed decay rate of binary systems PSR
B1913+16 \cite{key-1} and PSR J0348+0432 \cite{key-20}, one can
obtain some constraints on the strength of the $R^{2}$-dependent
term \cite{key-19,key-21,key-22}. There are many astrophysical situations
where the hot plasma of ionized atoms emits both electromagnetic waves
and GWs through the coulomb collisions between the electrons and ions
{[}23-26{]}. Hence, the analysis of the gravitational luminosity of
a plasma should be of general interest and may be, in principle, another
test for the validity of $f(R)$ thories of gravity. An expression
for the amount of radiated energy in a classical gravitational bremsstrahlung
in $R^{2}$-gravity has been derived in \cite{key-27} through the
assumption of the small-angle scattering approximation. In this paper,
we applied it to derive the gravitational luminosity of a hot plasma
with the gravitational bremsstrahlung as a mechanism for the energy
loss. After linearizing the $R^{2}$-gravity theory, we briefly discussed
the quadrupole radiation in $R^{2}$-gravity and the energy loss due
to gravitational bremsstrahlung in a single Coulomb collision between
two charged particles. Then, we calculated the thermal gravitational
radiation of the hydrogen plasma. Finally, we illustrated the correction
with an application to the Sun. The presence of the massive term in
Eq. (28) is a characteristic of higher-order terms in $f(R)$-gravity.
Thus, the $R^{2}$-gravity theory include massive GW modes. Hence,
our results beside confirming the standard GTR, stimulate the validity
of $f(R)$-gravity. Until now there is not available data to confront
our result to the experiment and fix the parameter of the $R^{2}$-gravity
contribution. We also stress the possibility to generalize the calculations
in this paper for other modified gravity theories listed in \cite{key-11}. 

\section{Acknowledgments }

This work has been supported financially by the Research Institute
for Astronomy and Astrophysics of Maragha (RIAAM). 

The authors thanks an unknown referee for useful comments.

\section*{Appendix: computation of some elementary integrals}

The set of following integrals have been used in evaluating (48) 
\begin{eqnarray*}
\int_{-\infty}^{\infty}\frac{x^{2}dx}{(a^{2}x^{2}+b^{2})^{3}} & = & \frac{\pi}{8a^{3}b^{3}},\\
\int_{-\infty}^{\infty}\frac{dx}{(a^{2}x^{2}+b^{2})^{2}} & = & \frac{\pi}{2ab^{3}}\\
\int_{-\infty}^{\infty}\frac{x^{4}dx}{(a^{2}x^{2}+b^{2})^{5}} & = & \frac{3\pi}{128a^{5}b^{5}},\\
\int_{-\infty}^{\infty}\frac{x^{2}dx}{(a^{2}x^{2}+b^{2})^{\frac{9}{2}}} & = & \frac{16}{105a^{3}b^{6}}\\
\int_{-\infty}^{\infty}\frac{x^{2}dx}{(a^{2}x^{2}+b^{2})^{4}} & = & \frac{\pi}{16a^{3}b^{5}},\\
\int_{-\infty}^{\infty}\frac{dx}{(a^{2}x^{2}+b^{2})^{4}} & = & \frac{5\pi}{16ab^{7}}\\
\int_{-\infty}^{\infty}\frac{dx}{(a^{2}x^{2}+b^{2})^{\frac{7}{2}}} & = & \frac{16}{15ab^{6}},\\
\int_{-\infty}^{\infty}\frac{x^{6}dx}{(a^{2}x^{2}+b^{2})^{6}} & = & \frac{3\pi}{256a^{7}b^{5}},\\
\int_{-\infty}^{\infty}\frac{x^{4}dx}{(a^{2}x^{2}+b^{2})^{6}} & = & \frac{3\pi}{256a^{5}b^{7}},\\
\int_{-\infty}^{\infty}\frac{x^{2}dx}{(a^{2}x^{2}+b^{2})^{5}} & = & \frac{5\pi}{128a^{3}b^{7}},\\
\int_{-\infty}^{\infty}\frac{dx}{(a^{2}x^{2}+b^{2})^{3}} & = & \frac{3\pi}{8ab^{5}}.
\end{eqnarray*}


\begin{thebibliography}{10}
\bibitem{key-1}R. A. Hulse and J.H. Taylor, Astrophys. J. \textbf{195},
L51 (1975).

\bibitem{key-2}B. Abbott et al. (LIGO Scientific Collaboration and
Virgo Collaboration), Phys. Rev. Lett. \textbf{116}, 061102 (2016).

\bibitem{key-3}A. Einstein, Sitzungsber. K. Preuss. Akad. Wiss. \textbf{1},
688 (1916).

\bibitem[4]{key-4}B. P. Abbott et al. (LIGO Scientific Collaboration
and Virgo Collaboration), Phys. Rev. Lett. \textbf{116}, 241103 (2016).

\bibitem[5]{key-5}F. Acernese et al., Journ. Phys. Conf. Ser. \textbf{32
(1)}, 223 (2006).

\bibitem[6]{key-6}F. Beauville et al., Class. Quant. Grav. \textbf{21
(5)}, S935 (2004).

\bibitem[7]{key-7}C. Corda, Int. J. Mod. Phys. D, 18, 2275 (2009).

\bibitem[8]{key-8}A. Einstein, Sitzungsber. K. Preuss. Akad. Wiss.\textit{\emph{
778 (1915).}}

\bibitem[9]{key-9}L. Landau L and E. Lifsits\textit{, Classical Theory
of Fields} (3rd ed.). London: Pergamon. ISBN 0-08-016019-0. Vol. 2
of the Course of Theoretical Physics (1971).

\selectlanguage{italian}%
\bibitem[10]{key-10}\foreignlanguage{english}{A. De Felice and S.
Tsujikawa, Living Rev. Rel. \textbf{13}, 3 (2010).}

\bibitem[11]{key-11}S. Nojiri, S. D. Odintsov, Phys. Rept. \textbf{505},
59 (2011).

\bibitem[12]{key-12}\foreignlanguage{english}{A. Pais, \textit{Subtle
Is the Lord: The Science and the Life of Albert Einstein''} - Oxford
University Press (2005).}

\bibitem[13]{key-13}\foreignlanguage{english}{G. 't Hooft, \textit{``The
mathematical basis for deterministic quantum mechanics''} - quant-ph/06/04/008
(2006).}

\bibitem[14]{key-14}P. J. E. Peebles\emph{, Principles of Physical
Cosmology}, Princeton University Press, Princeton (1993).

\selectlanguage{english}%
\bibitem[15]{key-15}\foreignlanguage{italian}{D. H. Lyth and A. R.
Liddle, \emph{Primordial Density Perturbation}, Cambridge University
Press (2009).}

\bibitem[16]{key-16}A. A. Starobinsky, Phys. Lett. B \textbf{91},
99 (1980).

\bibitem[17]{key-17}P. J. E. Peebles and B. Ratra, Rev. Mod. Phys.
\textbf{75,} 8559 (2003).

\bibitem[18]{key-18}\foreignlanguage{italian}{C. W. Misner, K. S.
Thorne and J. A. Wheeler, \emph{Gravitation}, W.H.Feeman and Company
(1973).}

\bibitem[19]{key-19}M. De Laurientis, S. Capozziello, Astropart.
Phys. \textbf{35}, 5, 257 (2011).

\bibitem[20]{key-20}J. Antoniadis et al., Science \textbf{340}, 6131
(2013).

\bibitem[21]{key-21}M. De Laurentis, I. De Martino, MNRAS \textbf{431
(1)}, 741 (2013).

\bibitem[22]{key-22}M. De Laurentis, I. De Martino, Int. Journ. Geom.
Meth. Mod. Phys. \textbf{12 (04)}, 1550040 (2015)

\bibitem[23]{key-23}S. Weinberg, Phys. Rev. \textbf{140}, B516 (1965). 

\bibitem[24]{key-24}G. Papini, S. R. Valluri, Phys. Rep. \textbf{33},
51 (1977). 

\bibitem[25]{key-25}F. H. Shu, \emph{The Physics Of Astrophysics},
University Science Books, Mill Valley, (1991). 

\bibitem[26]{key-26}G. B. Rybicki, A. P. Lightman, \emph{Radiative
Processes In Astrophysics}, John Wiley \textbackslash{}\& Sons, New
York (1979).

\bibitem[27]{key-27}A. Ajabshirizadeh, A. Jahan, B. Nadiri Niri,
Mod. Phys. Lett. A \textbf{29 (28)}, 1450145 (2014).

\bibitem[28]{key-28}G. F. R. Ellis, J. Murugan, C. G. Tsagas, Class.
Quant. Grav. \textbf{21}, 233-250 (2004).

\bibitem[29]{key-29}S. Nojiri and S. D. Odintsov,\foreignlanguage{italian}{
Phys. Rev. D \textbf{68}, 123512 (2003).}

\bibitem[30]{key-30}\foreignlanguage{italian}{C. Corda, Gen. Rel.
Grav. \textbf{40}, 10, 2201-2212 (2008).}

\selectlanguage{italian}%
\bibitem[31]{key-31}\foreignlanguage{english}{R. Jain, B. G. Sidharth,
C. Corda, Adv. High. En. Phys. 2601741 (2016).}

\bibitem[32]{key-32}C. Corda, H. J. Mosquera Cuesta, R.Lorduy Gòmez,
Astropart. Phys. \textbf{35}, 362 (2012).

\bibitem[33]{key-33}\foreignlanguage{english}{S. Nojiri and S.D.
Odintsov, Int. J. Geom. Meth. Mod. Phys. \textbf{4}, 115-146 (2007). }

\selectlanguage{english}%
\bibitem[34]{key-34}M. C. B. Abdalla et al.\textit{, THE PROBLEMS
OF MODERN COSMOLOGY, A volume in honour of Professor S.D. Odintsov
in the occasion of his 50th birthday - }Editor P.M. Lavrov, Copyright@2009
by Tomsk State Pedagogical University.

\bibitem[35]{key-35}C. D. Hoyle et al., Phys. Rev. Lett. \textbf{86},
4118 (2001).

\bibitem[36]{key-36}A. Accioly, S. Ragusa, E. C. de Rey Neto and
H. Mukai, Nuovo Cimento B \textbf{114}, 595 (1999).

\bibitem[37]{key-37}S. Capozziello, C. Corda, M. De Laurentis, Phys.
Lett. B \textbf{669 (5)}, 255 (2008).

\bibitem[38]{key-38}L. Lorenz, Phil. Mag. 34, 287 (1867).

\bibitem[39]{key-39}C. Corda, Phys. Rev. D \textbf{83}, 062002 (2011).

\bibitem[40]{key-40}H. Dehnen, F. Ghaboussi, Il Nuovo Cimento B \textbf{2,}
131 (1985) . 

\bibitem[41]{key-41}S. Capozziello, M. De Laurentis, S. Nojiri, S.
D. Odintsov, Gen. Rel. Grav. 41, 2313 (2009).

\bibitem[42]{key-42}C. Corda, New Adv. Phys. 7(1), 67 (2013).
\end{thebibliography}
\end{document}